\documentclass[twocolumn]{aastex702}

\received{June 1, 2020}
\revised{\today}
\accepted{\today}
\submitjournal{ApJL}

\shorttitle{WISE View of LyC Leakers}
\shortauthors{Yuan et al.}

\graphicspath{{./}{figures/}}

\def\h{~$h^{-1}$}
\def\m{~$\mu$m}
\def\R25{$R_{25}$}

\begin{document}

\title{Diversity in Lyman Continuum Escape at $z\sim0.3$ Revealed by WISE Infrared Observations}

\correspondingauthor{Fang-Ting Yuan}
\email{yuanft@shao.ac.cn}

\author{Fang-Ting Yuan}
\affiliation{Shanghai Astronomical Observatory, Chinese Academy of Sciences, 80 Nandan Road, Shanghai 200030, China}
\email[show]{yuanft@shao.ac.cn}

\author{Zhen-Ya Zheng}
\affiliation{Shanghai Astronomical Observatory, Chinese Academy of Sciences, 80 Nandan Road, Shanghai 200030, China}
\email{zhengzy@shao.ac.cn}

\author{Chunyan Jiang}
\affiliation{Shanghai Astronomical Observatory, Chinese Academy of Sciences, 80 Nandan Road, Shanghai 200030, China}
\email{cyjiang@shao.ac.cn}

\author{Shuairu Zhu}
\affiliation{Shanghai Astronomical Observatory, Chinese Academy of Sciences, 80 Nandan Road, Shanghai 200030, China}
\affiliation{University of Chinese Academy of Sciences, No. 19A Yuquan
Road, Beijing 100049, China}
\email{shuairuz@shao.ac.cn}

\begin{abstract}

The escape of Lyman continuum (LyC) radiation from star-forming galaxies plays a key role in cosmic reionization. While strong LyC leakers are commonly identified through ultraviolet (UV) and optical diagnostics, their infrared (IR) emission remains poorly explored. We use data from the Wide-field Infrared Survey Explorer (WISE) to investigate a sample of local star-forming galaxies, which contains 20 strong LyC leakers ($S/N_\mathrm{LyC} > 3$ and $f_{\rm esc}>5\%$) and 69 non-leakers. Among the strong leakers, 8 are classified with mid-IR detections. Comparing the IR-detected and IR-undetected subsamples, we find that the IR-undetected strong leakers exhibit higher [\ion{O}{3}]5007/[\ion{O}{2}]3726,3729 (O32) ratios, bluer UV slopes, and lower metallicities than the other subsamples. In contrast, the IR-detected strong leakers show O32 ratios, UV slopes, and metallicities comparable to those of non-leakers, while maintaining a median escape fraction of $f_{\rm esc}\sim12\%$. These results indicate that significant LyC escape is not limited to galaxies with the most extreme UV and optical properties and can coexist with substantial IR emission. The Ly$\alpha$ profiles and the morphology of the IR-detected and IR-undetected strong leakers imply that LyC photon escape in these two classes may be driven by different mechanisms. Our results highlight the diversity of LyC leakers and suggest that dusty star-forming galaxies may contribute a considerable amount to the ionizing photon budget during cosmic reionization.

\end{abstract}


\section{Introduction} 
\label{sec:intro}

Identifying the physical mechanisms that enable Lyman Continuum (LyC) radiation to escape from galaxies is essential for understanding their contribution to the Epoch of Reionization (EoR).
Due to the opacity of the intergalactic medium at $z > 6$, researchers rely on samples at lower redshifts ($z<4.5$) to identify empirical diagnostics for LyC leakage \citep[e.g.,][]{bergvall2006,debarros2016,vanzella2016,Izotov2016a,saha2020,yuan2021,pahl2021,pahl2023,begley2022,Flury2022}. In particular, local LyC emitters (LCEs or leakers) at $z \sim 0.3$ provide a unique advantage. At these low redshifts, the impact of IGM absorption is negligible, enabling a robust determination of the absolute LyC escape fraction ($f_{\mathrm{esc}}$). 

However, whether these local analogs truly represent the high-redshift population remains unclear. Recent work suggests that high-redshift LyC leakers may be characterized by two distinct physical modes: extreme starbursts and merger-driven systems \citep{yuan2024, zhu2025, lereste2025merger}. This dichotomy highlights the potential diversity in the interstellar medium (ISM) geometries of leakers, which may not be fully captured by traditional low-redshift selection functions.

Furthermore, although early investigations of LyC escape often include infrared‑bright systems such as Luminous and ultraluminous infrared galaxies \citep[LIRGs and ULIRGs; e.g.,][]{leitherer1995,Heckman2001,deharveng2001,Grimes2009,cormier2012}, more recent work has shifted toward low‑mass and dust‑poor galaxies, which are now commonly used as local analogs of LyC leakers. As a result, the infrared (IR) properties of known LyC leakers have received relatively little attention. In fact, despite low-to-moderate $f_\mathrm{esc}$, LyC leakage has been detected in several IR‑active galaxies \citep[e.g.,][]{bergvall2006,leitet2013,borthakur2014,roy2025}, indicating that the connection between dust obscuration and LyC escape remains uncertain. 

Recent studies have highlighted the significance of dust-obscured star formation even at the EoR, suggesting that dust may have accumulated much earlier than previously assumed \citep[e.g.,][]{watson2015, laporte2017,burgarella2020,fudamoto2021,alvarez-marquez2023, sun2026}. On the other hand, faint galaxies dominate the number counts of the rest-UV luminosity function in the EoR \citep[e.g.,][]{finkelstein2019}, and at lower redshifts, there is evidence that $f_\mathrm{esc}$ is higher in UV-fainter, lower-mass, and possibly bluer galaxies \citep[e.g.,][]{chisholm2022,pahl2023,jaskot2024,mascia2025}. LyC leakers are thus expected to be predominant during the EoR. Therefore, we find it imperative to systematically investigate their IR properties to reconcile their dust content with their ionizing escape fractions.

In this work, we use the {\it Wide-field Infrared Explorer} \citep[WISE;][]{Wright2010} archival data to study a sample of LyC leakers at $z \sim 0.3$. We aim to investigate whether IR observables can provide new insights into the ISM conditions of leakers, particularly in identifying the porous geometries that might allow LyC photons to escape from seemingly dusty or massive systems.
We describe the sample and data in Section \ref{sec:sample_and_data}. The main results are presented in Section \ref{sec:results}, followed by a discussion of the LyC escape pathways in strong leakers in Section \ref{sec:discussion}. A summary of our findings is provided in Section \ref{sec:summary}.

Throughout, we use Planck cosmology \citep{planck2016}, with $H_0 = 66.9$ km s$^{-1}$ Mpc$^{-1}$, $\Omega_m = 0.32$, and $\Omega_\Lambda = 0.68$.

\section{Sample and Data}
\label{sec:sample_and_data}
\subsection{Sample of Low-Redshift LCEs}
We adopt the homogeneous galaxy sample of 89 star-forming galaxies at z $\sim$ 0.3 compiled by \citet{Flury2022} from HST/COS observations. The sample comprises 66 Low-z Lyman Continuum Survey (LzLCS) galaxies (35 confirmed leakers) selected from SDSS star-forming galaxies with high-excitation, high star formation surface density, or blue UV slopes, together with 23 galaxies collected from previous studies \citep[15 confirmed leakers,][]{Izotov2016a, Izotov2016b, Izotov2018a, Izotov2018b, Wang2019}. The sample spans a broad range of physical parameters, including the [\ion{O}{3}]5007/[\ion{O}{2}]{3726,3729} (hereafter O32) line ratio, UV continuum slope (\(\beta_{1200}\)), stellar mass, star formation rate (SFR), metallicity, and burst age. LyC fluxes and escape fractions were measured consistently across the sample.

The sample can be divided into two groups based on their LyC detections. The strong LyC leakers, following the nomenclature of \citet{flury2022b} and \citet{saldana-lopez2022}, are sources with S/N $>5$ and $f_\mathrm{esc} \geq 0.05$. These sources have significant signals of LyC emissions and considerable escape fractions of ionizing photons. The remaining galaxies are referred to as non-leakers, as their LyC detections are not as robust as those of strong leakers. The sample consists of 20 strong leakers and 69 non-leakers.

\subsection{Infrared Data}

We investigate the IR properties of the sample based on the WISE observations. The angular resolutions of the W1–W4 bands (corresponding to 3.4, 4.6, 12, and 22 $\mu$m) are approximately 6.1\arcsec, 6.4\arcsec, 6.5\arcsec, and 12.0\arcsec, respectively, with 5$\sigma$ point-source sensitivities in unconfused regions of about 0.08, 0.11, 1, and 6 mJy (\citealt{Wright2010}; \citealt{Cutri2013}).

We cross-match the sample with the AllWISE catalog \citep{allwise2019} using a $3''$ radius. We then inspect the WISE images and remove 4 sources whose fluxes are contaminated by nearby sources. To investigate the dust and SFR-related properties, we examine whether a source is detected at $>2\sigma$ in the W3 or W4 bands. At $z\sim0.3$, the two bands probe rest-frame $\sim9\,\mu$m and $\sim17\,\mu$m, respectively, dominated by aromatic band emission from polycyclic aromatic hydrocarbons (PAHs) and continuum emission from warm dust. Emissions in these wavelengths have been shown to correlate with total IR luminosity and star formation activity in star-forming galaxies \citep[e.g.,][]{wu2005,takeuchi2005,yuan2012,salim2016,cluver2017}. On the other hand, the W1 and W2 bands contain substantial stellar continuum contributions \citep[e.g.,][]{leroy2019}. We find that 31 sources are detected in the W3 or W4 band, and 54 sources are not detected. Hereafter, we refer to them as IR-detected and IR-undetected subsamples, respectively.

To further compare the properties of the IR-detected and IR-undetected subsamples, we perform a stacking analysis of the W3-band images for the IR-undetected subsamples and calculate the typical luminosities $L_\mathrm{w3} \equiv \nu L_{\nu}(W3)$ of them from the stacked images. We take an $88\times88$ pixel ($2'\times2'$) cutout centered around each source, and then stack these subimages to calculate a mean flux density at each pixel position. We measured the flux of the stacked image using aperture photometry with a radius of 13.75\arcsec (5 pixels). At larger radii, the enclosed W3 flux becomes approximately stable. The uncertainties of the fluxes are determined using bootstrap resampling. The final flux is converted to luminosity using the average redshift.

\subsection{Measurements of Physical Properties}
\label{subsec:phys_prop}

To analyze the physical properties of different subsamples, we adopt the measurements from \citet{Flury2022, flury2022b}, including \(f_{\mathrm{esc}}\), SFR$_{\mathrm{H}\beta}$, age, stellar masses ($M_*$), dust extinction ($E(B-V)$), half-light radius based on the COS images ($r_{50}^\mathrm{cos}$), the equivalent width of Ly$\alpha$ EW(Ly$\alpha$), and the UV continuum slope $\beta_{1200}$ derived from COS spectra.

We independently measure O32 by integrating the emission-line fluxes using the spectra from the Sloan Digital Sky Survey (SDSS). We measure [\ion{O}{3}]5007 and [\ion{O}{2}]3726,3729 fluxes by directly integrating continuum-subtracted SDSS spectra within $\pm$18\,\AA\ windows. This window size is sufficiently wide to encompass the full line profiles, and we have tested that varying the window to $\pm15$\AA{} or $\pm20$ \AA does not introduce significant differences in the measured fluxes. The continuum is estimated from adjacent line-free regions. All fluxes are corrected for dust attenuation using $E(B-V)$ provided in \citet{Flury2022} and the CCM extinction law \citep{cardelli1989}. Similar to \citet{Flury2022}, when the [\ion{O}{3}]5007 line is affected by saturation or sky lines, we replace its flux $F_{5007}$ with 2.98$F_{4959}$, assuming the theoretical [\ion{O}{3}]5007/[\ion{O}{3}]4959 ratio of 2.98 \citep{storeyzeippen2000}. 
We also fit the emission lines with Gaussian profiles, using a single Gaussian for [\ion{O}{3}]5007 or [\ion{O}{3}]4959 and a double-Gaussian model for the \ion{O}{2} doublet. The resulting O32 values agree well with those obtained from direct integration, showing a median difference of only 0.002 dex and a scatter of 0.02 dex.

We further compare our O32 measurements with those reported by \citet{Flury2022}. Our measurements are systematically lower by $\sim0.15$ dex, with a scatter of 0.16 dex. This offset likely reflects differences in continuum treatment, as \citet{Flury2022} fit global continua with polynomial functions, whereas we adopt constant local continuum levels. Despite this difference, we have verified that our scientific conclusions remain unchanged when using either set of O32 measurements. In the following analysis, we adopt the direct-integration values.

We use Code Investigating GALaxy Emission \citep[CIGALE][]{burgarella2005,noll2009,boquien2019} to fit the UV-to-IR SEDs of these galaxies using GALEX, SDSS, and available WISE photometry, with the WISE data providing additional constraints on the dust emission. We assume a Chabrier initial mass function \citep[IMF;][]{chabrier2003} and adopt the stellar population models of \citet{bc2003} (BC03). The star formation history (SFH) is assumed to follow a double-exponential form (\texttt{sfh2exp}), as defined in Equation (1) of \citet{boquien2019}. Dust attenuation is described using the \citet{calzetti2000} law with a variable slope modification \citep{noll2009}, while dust emission is modeled using the templates of \citet{dl2014}. The adopted dust emission models are based on a dust mixture of amorphous silicate and graphite grains, and PAH. Combined with the dust attenuation model, the dust parameters (mass $M_\mathrm{dust}$ and luminosity $L_\mathrm{dust}$) are derived from the SED fitting under the energy-balance assumption implemented in CIGALE, where the stellar energy absorbed by dust is re-emitted in the infrared. The global quality of the fit is evaluated using the reduced $\chi^2$ of the best-fit model, and physical parameters are derived following the Bayesian approach implemented in CIGALE.

\section{Results}
\label{sec:results}
\subsection{IR-bright and IR-faint systems}

\begin{table}
\centering
\caption{WISE W3/W4 detection statistics of the sample.}
\label{tab:detrate}
\begin{tabular}{lcccc}
\hline\hline
Category & IR-det. & IR-undet. & Total & IR Fraction \\
\hline
Strong Leaker    & 8 &  12 & 20 & 40\% \\
Non-leaker & 23 & 42 & 65 & 35\% \\
All        & 31 & 54 & 85 & 36\% \\
Strong Fraction & 26\% & 22\% & 24\% & \\
\hline
\end{tabular}
\end{table}

\begin{table*}
\centering
\footnotesize
\setlength{\tabcolsep}{3pt}
\caption{IR-detected strong LyC leakers identified from the WISE W3/W4 bands.}
\label{tab:irdetected_lce}
\begin{tabular}{lccccccc}
\toprule

name & z & $f_\mathrm{esc}$ (\%) & log$M_* (M_{\odot})$ & S/N(W3) & $\log(L_\mathrm{w3}/L_{\odot})$ & S/N(W4)  & $\log(L_\mathrm{w4}/L_{\odot})$\\
\hline
J103344+635317 & 0.35 & $30.5^{+15.4}_{-11.6}$ & $9.13\pm0.43$ & $\cdots$ & $\cdots$ &3.92 & $10.73\pm0.11$ \\
J1152+3400 & 0.34 & $17.7^{+19.5}_{-6.3}$ & $9.59\pm0.15$ & 2.00 & $9.84\pm0.22$ & $\cdots$ & $\cdots$ \\
J091703+315221 & 0.30 & $16.1^{+7.3}_{-5.5}$ & $9.31\pm0.43$ & 2.57 & $9.89\pm0.17$ &2.26 & $10.44\pm0.19$ \\
J1333+6246 & 0.32 & $12.8^{+12.6}_{-6.2}$ & $8.50\pm0.13$ & 2.04 & $9.51\pm0.21$ & $\cdots$ & $\cdots$ \\
J1442-0209 & 0.29 & $12.0^{+6.4}_{-5.0}$ & $8.96\pm0.14$ & 5.84 & $10.09\pm0.07$ &5.05 & $10.58\pm0.09$ \\
J141013+434435 & 0.36 & $10.5^{+7.1}_{-7.0}$ & $8.72\pm0.43$ & 6.62 & $10.35\pm0.07$ &5.54 & $10.81\pm0.08$ \\
J115855+312559 & 0.24 & $6.6^{+3.0}_{-1.5}$ & $9.75\pm0.35$ & 6.87 & $9.98\pm0.06$ &4.85 & $10.46\pm0.09$ \\
J143256+274249 & 0.27 & $6.0^{+0.9}_{-0.7}$ & $10.54\pm0.16$ & 22.62 & $10.54\pm0.02$ &13.57 & $10.98\pm0.03$ \\
\hline
\end{tabular}
\end{table*}

\begin{figure*}
\centering
    \includegraphics[width=0.95\linewidth]{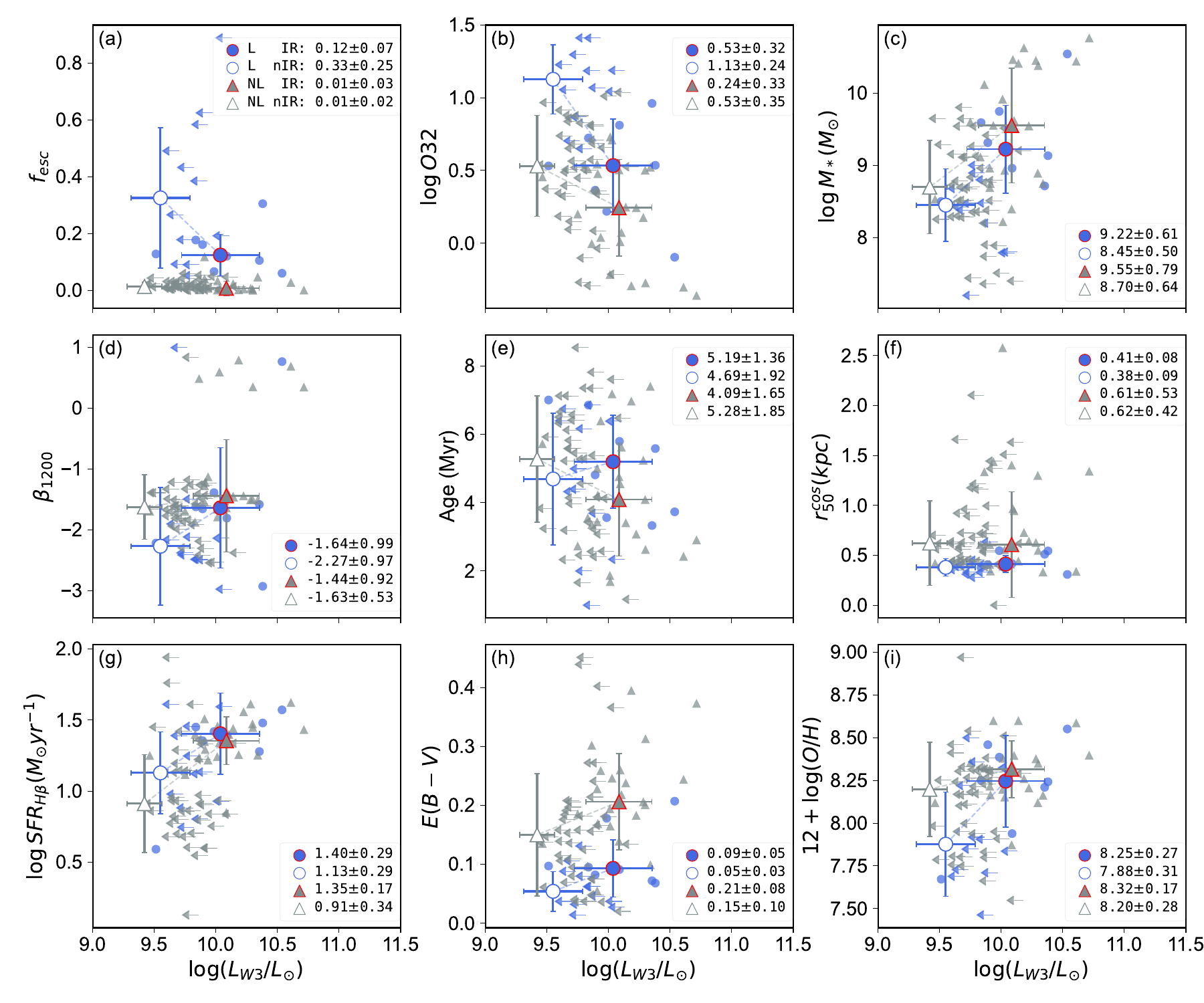}\\
    \caption{(a)–(i) show galaxy properties as a function of $L_{W3}=\nu L_{\nu}(W3)$. Strong and non-leaker samples are shown in blue circles and gray triangles, respectively. Individual IR-detected galaxies are shown as filled symbols, while IR-undetected galaxies are indicated by leftward arrows representing the $2\sigma$ upper limits. For each IR-undetected subsample, stacked measurements are shown as open symbols with error bars, while the medians of IR-detected sources are shown as filled symbols with red edges.}
    \label{fig:box_plot}
\end{figure*}

Based on the W3/W4-band selections, the sample can be divided into IR-bright (detected, 31 galaxies) and IR-faint (undetected, 54 galaxies) systems. {When combined with the LyC detections, this yields four subsamples,} summarized in Table \ref{tab:detrate}. The eight IR-detected strong LyC leakers are listed in Table \ref{tab:irdetected_lce}, together with their basic properties and W3/W4 detection significances.

The fraction of IR detections in strong LyC leakers ($\sim 40\%$) and non-leakers ($\sim 35\%$) is similar. Conversely, the fractions of strong LyC leakers in the IR-detected and IR-undetected subsamples are also similar (26\% versus 22\%). The IR-detected strong leakers and non-leakers have W3 band luminosity $L_\mathrm{w3}$ about $10^{10}L_{\odot}$. The IR-undetected sources are fainter: The stacking analysis shows that the typical luminosity is $\log L_\mathrm{w3}=9.55\pm0.04$ for the IR-undetected strong leakers and $\log L_\mathrm{w3}=9.42\pm0.04$ for non-leakers.

The presence of a substantial fraction of LyC leakers in the IR-detected subsample suggests that significant LyC escape is still possible in relatively dust-rich environments. This is consistent with previous findings that some IR-bright Lyman Break Analogs may leak a considerable fraction of ionizing radiation (e.g., \citealt{heckman2011, borthakur2014}).

The $f_\mathrm{esc}$ of LyC photons in each subsample is shown in Figure \ref{fig:box_plot} (a). Within the strong leakers, IR-undetected sources exhibit systematically higher escape fractions than IR-detected ones (median $f_\mathrm{esc} = 32.6\pm24.7\%$ vs. $12.4\pm7.3\%$). Despite this internal difference, IR-detected strong leakers still maintain $f_\mathrm{esc}$ values significantly exceeding those of the non-leakers, whose $f_\mathrm{esc}$ remains consistently low (median $\sim 1\%$) regardless of IR detection.

We examine the potential overestimation of $f_\mathrm{esc}$ when SFRs are derived without IR constraints. Using SFRs from our SED fitting that includes IR data, the median $f_\mathrm{esc}$ for IR-detected strong leakers decreases to \textbf{$9.7 \pm 8.2\%$}. 
This $f_\mathrm{esc}$ still remains significantly higher than that of non-leakers.

\subsection{Infrared detection versus O32}

\begin{figure}
    \centering
    \includegraphics[width=\linewidth]{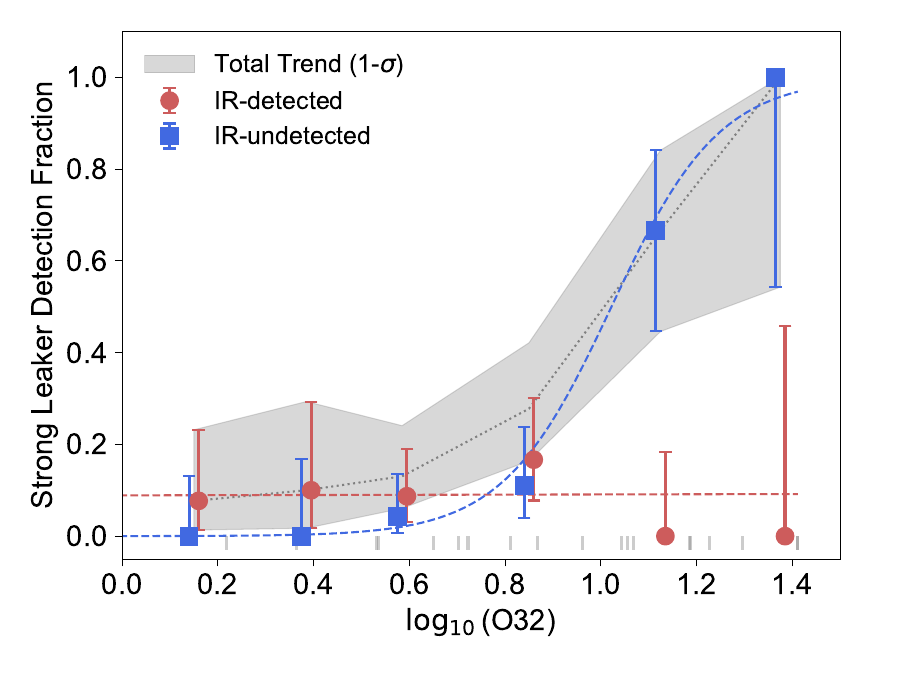}
    \caption{LyC detection fraction as a function of $\log$O32. The IR‑undetected systems (blue squares) show a rising LyC detection fraction with increasing O32 (blue dashed line), consistent with O32 tracing globally high ionization conditions. In contrast, the IR‑detected strong leakers (red dots) exhibit an approximately constant detection fraction across the full O32 range (red dashed line), indicating that their LyC escape is not strongly linked to the galaxy‑wide O32 ratio.  The shaded region indicates the overall detection fraction for the combined sample. At low O32, the total detection fraction is dominated by the IR‑detected population, producing a weak trend in the combined sample.}
    \label{fig:strong_detfrac}
\end{figure}

We assess the ISM ionization state using O32 for both the IR‑detected and IR‑undetected systems. Within the strong LyC leakers, the two subsamples exhibit a clear and systematic difference in O32. IR-undetected strong leakers preferentially exhibit higher O32 values, whereas IR-detected strong leakers occupy a regime of comparatively lower O32 (Figure \ref{fig:box_plot} (b)). The median O32 of the IR-undetected strong leakers is higher by $\Delta$O32 $\approx10.11\pm2.28$ than that of IR-detected strong leakers (Table \ref{tab:IR_comparison}). We assess the significance using a permutation test \citep[e.g.,][]{good2005}. Specifically, we randomly shuffle all data points in the two subsamples 50000 times and recompute the median difference for each permutation. The $p$-value is estimated as the fraction of permutations with a median difference greater than the observed one. The test yields a $p$-value of $0.01$, indicating that the observed difference is unlikely to arise by chance. A bootstrap analysis gives a consistent median difference of $-9.94\pm2.75$, with a 95\% confidence interval of (-15.20, -4.85). Both tests show that the difference between the two subsamples is statistically significant.

While our sample has been pre-screened via optical spectroscopy to exclude AGN contamination, obscured AGN activity remains possible. Using WISE color-color criteria \citep{Stern2012}, we identify one IR-detected strong leaker, J1442-0209, at the edge of the AGN selection region. This source exhibits the second highest O32 ratio ($\sim6.47\pm0.57$) among the IR-detected strong leakers. If we remove this source from our sample, the difference between the two subsamples would be further enhanced.

This systematic difference in the ionizing state between IR-detected and IR-undetected strong leakers may suggest that the two subsamples reflect two distinct LyC-leaking regimes.
The IR‑undetected strong leakers exhibit globally elevated ionization conditions consistent with a density‑bound geometry. On the other hand, the IR‑detected strong leakers show low O32 despite their substantial LyC leakage. These strong leakers present O32 similar to non-leakers, whether IR-detected or not.

It seems that O32 does not trace the high escape fractions in the IR-detected strong leakers, although previous studies consider high O32 to be a necessary or primary indicator of LyC escape \citep{nakajima2020,jaskot2024}. One possible interpretation is that the IR‑detected strong leakers likely allow LyC photons to escape through a limited number of low‑density channels rather than through a galaxy‑wide density‑bounded medium \citep[e.g.,][]{gazagnes2018,saldana-lopez2022}. In such a “picket‑fence’’ or aperture‑driven leakage scenario, the global O32 ratio does not necessarily reflect the local conditions along the escaping sightlines. As a result, even galaxies with relatively modest O32 values can still exhibit high LyC escape fractions if the escape occurs through narrow, low‑opacity pathways. In contrast, for the IR‑undetected strong leakers, O32 remains an effective indicator of LyC escape: these galaxies exhibit globally elevated ionization conditions consistent with density‑bounded geometries. We further discuss the possibility of distinct LyC-leaking pathways in Section~\ref{subsec:twopops}.

This distinction can explain why O32 loses its robustness in tracing LyC leakage in some cases, which is not expected in a density-bound scenario. As shown in Figure \ref{fig:strong_detfrac}, the detection fraction increases monotonically with O32 for the IR-undetected strong LyC leakers. However, this trend does not hold for the IR‑bright strong leakers: their detection rate remains almost flat as the O32 increases. 

\subsection{Comparison of Physical Properties of IR-detected and IR-undetected strong LyC leakers}
\label{subsec:physprop}
\begin{table*}[t]
\centering
\caption{Median physical properties of the IR‑detected and IR‑undetected subsamples. The number of galaxies in each subsample is given in parentheses.}
\label{tab:IR_comparison}
\begin{tabular}{lcccc}
\hline\hline
 & \multicolumn{2}{c}{Strong Leakers} 
 & \multicolumn{2}{c}{Non--leakers} \\
Property 
 & IR--non (12) & IR--det (8) 
 & IR--non (42) & IR--det (23)\\
 
\hline
$f_\mathrm{esc}$ & $0.33 \pm 0.25$ & $0.12 \pm 0.07$ & $0.01 \pm 0.02$ & $0.01 \pm 0.03$ \\
$\log O32$ & $1.13 \pm 0.24$ & $0.53 \pm 0.32$ & $0.53 \pm 0.35$ & $0.24 \pm 0.33$ \\
$\log M_* (M_{\odot})$ & $8.45 \pm 0.50$ & $9.22 \pm 0.61$ & $8.70 \pm 0.64$ & $9.55 \pm 0.79$ \\
$\beta_{1200}$ & $-2.27 \pm 0.97$ & $-1.64 \pm 0.99$ & $-1.63 \pm 0.53$ & $-1.44 \pm 0.92$ \\
Age (Myr) & $4.69 \pm 1.92$ & $5.19 \pm 1.36$ & $5.28 \pm 1.85$ & $4.09 \pm 1.65$ \\
$r_{50}^\mathrm{cos}~\mathrm{(kpc)}$ & $0.38 \pm 0.09$ & $0.41 \pm 0.08$ & $0.62 \pm 0.42$ & $0.61 \pm 0.53$ \\
$\log \mathrm{SFR_{H\beta}} (M_{\odot} yr^{-1})$ & $1.13 \pm 0.29$ & $1.40 \pm 0.29$ & $0.91 \pm 0.34$ & $1.35 \pm 0.17$ \\
$E(B-V)$ & $0.05 \pm 0.03$ & $0.09 \pm 0.05$ & $0.15 \pm 0.10$ & $0.21 \pm 0.08$ \\

EW(Ly$\alpha$) ($\AA$) & 127.00$\pm$60.63 & 67.62$\pm$36.86 &
49.59$\pm$58.07 & 32.15$\pm$30.90 \\

$\log L_{W3}~(L_{\odot})$ & $9.55 \pm 0.24$ (stack) & $10.04 \pm 0.32^b$ & $9.42 \pm 0.14$ (stack) & $10.08 \pm 0.27^b$\\

{$\log L_\mathrm{dust} (L_\odot)^a$} & $10.46\pm0.29$ & $11.17\pm0.32$ & $10.69\pm0.35$ & $11.35\pm0.31$ \\ 

{$\log M_\mathrm{dust} (M_\odot)^a$} & $6.90\pm0.30$ & $7.60\pm 0.33$ & $7.14\pm0.38$ & $7.83\pm 0.38$ \\
\hline
\end{tabular}

\vspace{2pt} 
\begin{flushleft}
\small
$^a$ Dust luminosities and masses are derived using CIGALE SED fitting with the \citet{Draine2014} dust models, as described in Section \ref{subsec:phys_prop}.

{$^b$} For two galaxies detected in the W4 band but not in W3 (one in strong leakers, the other in non-leakers), we use the $L_{W3}$ values derived from the CIGALE best-fit model.

\end{flushleft}
\end{table*}

Table \ref{tab:IR_comparison} summarizes the median values and standard deviations of the physical properties of the IR-detected and IR-undetected subsamples. As expected, the IR-detected systems have higher dust luminosities ($L_\mathrm{dust}$) and dust masses ($M_\mathrm{dust}$) than IR-undetected systems. Figure \ref{fig:box_plot} shows the physical properties derived from UV and optical spectra as a function of $L_{W3}$ for all subsamples. The IR-undetected strong leakers exhibit the highest O32, the bluest $\beta_\mathrm{1200}$, and the lowest metallicity among all subsamples, even compared to the IR-undetected non-leakers with similar dust luminosity and stellar mass. These properties match the classical conditions often associated with efficient LyC escape.

In contrast, the IR-detected strong leakers exhibit O32, $\beta_{1200}$, and metallicity similar to non-leakers rather than IR-undetected strong leakers. The high stellar mass and red $\beta_{1200}$ of IR-detected strong leakers seem to deviate from the empirical trends established in previous studies \citep[e.g.][]{chisholm2022, pahl2023}. One possible explanation is that the IR-detected strong leakers simply represent the higher-mass end of the LyC-leaking population. However, although the IR-detected strong leakers tend to have higher stellar masses on average, the two leaker populations overlap at $\log(M_*/M_\odot)\sim8.5$--9.0, indicating that stellar mass alone cannot account for the infrared classification. Moreover, both leakers and non-leakers occupy a similar stellar mass range, suggesting that stellar mass alone is insufficient to distinguish LyC-leaking galaxies from non-leakers.

The most significant property that can distinguish the IR-detected strong leakers from non-leakers is the $E(B-V)$ derived from the optical spectra. Despite being IR-luminous, these strong leakers show significantly lower $E(B-V)$ values than non-leakers, suggesting a spatial decoupling between their IR and optical emissions. As in the ``picket-fence" configuration, the ionizing radiation escapes through low-density channels, avoiding the dense dust clouds that drive the IR luminosity. Conversely, the higher $E(B-V)$ observed in non-leakers indicates an ISM with a higher covering factor, where diffuse dust acts as a continuous screen that both increases optical reddening and suppresses LyC escape. This distinction may also be attributed to orientation effects \citep{maulick2026}: non-leakers may host similar channels, but they are not aligned with our sightline. In this sense, the difference between IR-detected strong leakers and non-leakers likely arises from the distribution and geometry of stars, gas, and dust, rather than the global galaxy properties (as further discussed in Section \ref{subsec:dust}). 

The Ly$\alpha$ equivalent width is correlated with the HI coverage fraction and is thus a statistical indicator of the LyC escape fraction \citep[e.g.,][]{steidel2018,gazagnes2020,Flury2022}. We compare the EW(Ly$\alpha$) across the subsamples (Table \ref{tab:IR_comparison}). The IR-undetected strong leakers show a markedly higher median EW(Ly$\alpha$) than the other subsamples, consistent with their higher median $f_\mathrm{esc}$. The IR-detected strong leakers also have a slightly higher median EW(Ly$\alpha$) than the non-leakers, though the difference is small. Their lower EW(Ly$\alpha$) compared to the IR-undetected leakers aligns with their relatively lower median $f_\mathrm{esc}$. We discuss the Ly$\alpha$ profiles of the strong leakers in more detail in Section \ref{subsec:twopops}.

\section{Discussion}
\label{sec:discussion}
\subsection{Multiple pathways for strong LyC escape}
\label{subsec:twopops}

\begin{figure*}
    \centering
    \includegraphics[width=\linewidth]{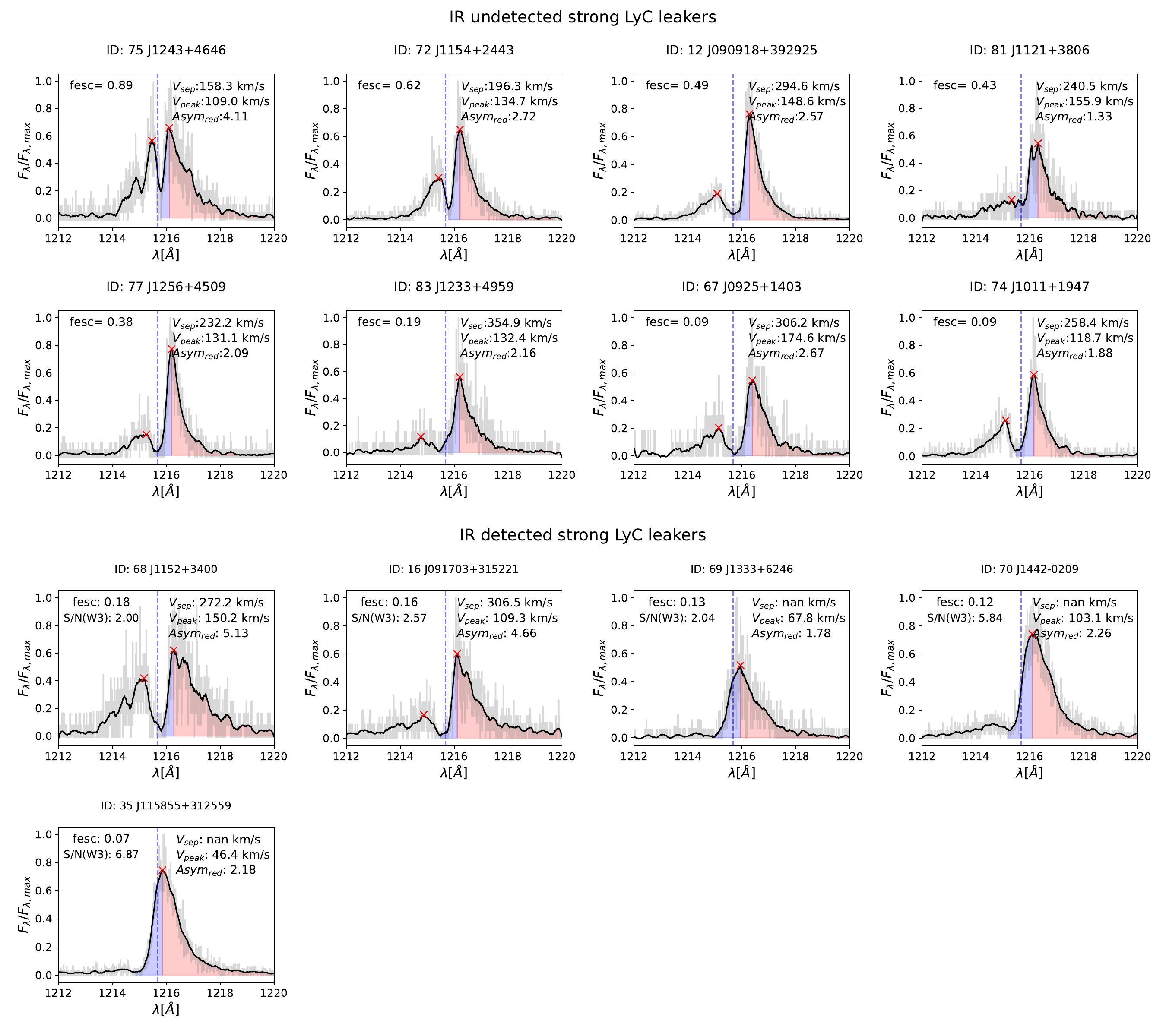}
    \caption{Ly$\alpha$ profiles of IR-detected and IR-undetected strong leakers observed with HST/COS G160M. Peaks used for the calculations are marked with crosses. We measure the velocity offset of the dominant peak relative to the systemic redshift ($v_{\mathrm{peak}}$) and the peak separation ($v_{\mathrm{sep}}$). The red-side asymmetry parameter ($A_r$; \citealt{kakiichi2021}) is computed as the flux ratio between the red and blue regions indicated in the figure, corresponding to the integration windows used in the analysis. The dashed line marks the rest-frame Ly$\alpha$ wavelength.}
    \label{fig:lya_profile}
\end{figure*}

Our analysis indicates that IR-detected and IR-undetected strong LyC leakers may represent different pathways of LyC escape. The IR-faint systems appear broadly consistent with a density-bound scenario. In these galaxies, a low neutral hydrogen column density likely allows the ionized region to extend to the galactic boundary. Given a standard dust-to-gas ratio, such systems would naturally be dust-poor, leading to faint, and therefore undetected, IR emission.

Conversely, the IR-bright systems may be interpreted as ionization-bound geometries with a porous ISM. In this framework, substantial dust reservoirs ($M_\mathrm{dust}\approx3.0\times10^7M_\odot$) could coexist with LyC leakage if the escape is highly anisotropic, occurring through low-column-density channels. While the ionizing photons are absorbed and reprocessed into detectable IR emission, a significant fraction can still escape through these channels. This would explain why their global observables (like O32 or $\beta_{1200}$) appear decoupled from their high $f_\mathrm{esc}$. These systems may share the same leakage geometry as the high‑redshift sources that exhibit spatially offset LyC emission relative to the rest of the galaxy \citep[e.g.,][]{yuan2024,gupta2024,maulick2025,ji2025,rivera-thorsen2025}.

The Ly$\alpha$ profile may provide information on the gas density and geometry of a galaxy \citep[e.g.,][]{schaerer2011,verhamme2015,hu2023,pahl2024}. In our sample, 8 IR-undetected and 5 IR-detected strong leakers were observed with the COS medium-resolution spectrograph (G160M). The corresponding Ly$\alpha$ profiles are shown in Figure \ref{fig:lya_profile}. We measure the velocity offset of the dominant peak ($v_{\mathrm{peak}}$), the peak separation ($v_{\mathrm{sep}}$), and the red-side asymmetry parameter ($A_r$; \citealt{kakiichi2021}) from these spectra. These parameters provide information on the distribution of gas and LyC-leaking conditions \citep{vanzella2015,kakiichi2021}. For the IR-undetected strong leakers, we find that 6 of these 8 sources satisfy $v_\mathrm{sep} < 300$ km/s, and 7 sources satisfy $v_\mathrm{peak} < 150$ km/s, indicating low H$\,\textsc{i}$ column densities and a high likelihood of LyC leakage \citep{verhamme2015}. J0925+1403 represents a borderline case; its $v_\mathrm{sep}$ ($306.2$ km/s) and $v_\mathrm{peak}$ ($174.6$ km/s) slightly exceed the threshold. Furthermore, 7 IR-undetected sources display a red peak asymmetry $A_r < 3$, suggesting that these systems may be closer to a density-bound regime where LyC photons escape through a globally low H$\,\textsc{i}$ column density.
The only exception is J1243+4646, whose higher asymmetry ($A_r = 4.11$) is likely associated with its complex multi-peaked profile, possibly reflecting a more disturbed gas geometry compared to the rest of the subsample.

Regarding the 5 IR-detected sources, 2 strong leakers present a multi-peaked structure where the two primary peaks fulfill the escape criteria ($v_\mathrm{peak} < 150$ km/s, or $v_\mathrm{sep}< 300$ km/s). Their high red-peak asymmetry ($A_r > 3$) supports a ``picket-fence" scenario where leakage occurs through low-density H$\,\textsc{i}$ holes in a clumpy ISM \citep{kakiichi2021,hu2023}. In contrast, 3 IR-detected strong leakers 
display a single dominant peak centered near the systemic redshift ($v_\mathrm{peak} <150$ km/s) with lower $A_r\sim2$, potentially indicating significant escape through a direct line-of-sight channel \citep{verhamme2015,hu2023}. We note that, within the current sample, single-peaked Ly$\alpha$ profiles are found only among the IR-detected sources. Based on the current analysis of the Ly$\alpha$ profiles, the IR-detected strong leakers appear to be more consistent with Ly$\alpha$ escape through low-density channels in a clumpy medium, while the IR-undetected strong leakers are more consistent with the high-$f_\mathrm{esc}$ density-bound dominated regime discussed by \citet{kakiichi2021}. 

To further characterize the ISM of these strong LyC leakers, we utilize 3 GHz and 6 GHz radio continuum data from the VLA observation presented by \citet{bait2024}. At these frequencies, the radio flux density provides a dust-extinction-free probe of dense star-forming regions, offering an independent diagnostic of the ISM conditions that complements our optical and IR observations. Although this radio analysis is limited to a subsample of 6 sources, the results reveal a striking 100\% correspondence between IR and radio detections that reinforces our proposed classification. Specifically, 2 IR-undetected leakers are also undetected in the VLA radio bands, suggesting a less substantial or lower-density ISM that lacks the gas density necessary to sustain robust radio emission. In contrast, 4 IR-bright strong leakers are detected at both 3 GHz and 6 GHz, indicating substantial gas content. In such a gas‑rich ISM, LyC photons can escape only if the medium is porous, with low‑density channels that allow them to pass through.

The differences between the IR-detected and IR-undetected strong leakers presented in Section \ref{subsec:physprop}, together with the Ly$\alpha$ and radio properties, suggest that the two subsamples may occupy different regimes of LyC escape. The IR-detected strong leakers have lower $f_{\rm esc}$ and substantial dust and gas, and are naturally consistent with porous escape through low-covering-fraction channels. This interpretation is in line with previous absorption-line studies emphasizing the importance of low HI covering fractions for LyC leakage \citep{gazagnes2018,chisholm2018,gazagnes2024,saldana-lopez2022}. These studies show that LyC photons can escape through low-column-density channels even when a substantial neutral gas reservoir is present. In particular, the inferred low covering fractions of strong leakers suggest that the covering fraction of neutral gas, rather than the global gas column density alone, can be a key parameter regulating the LyC escape fraction. 

\begin{figure}
    \centering
    \includegraphics[width=\linewidth]{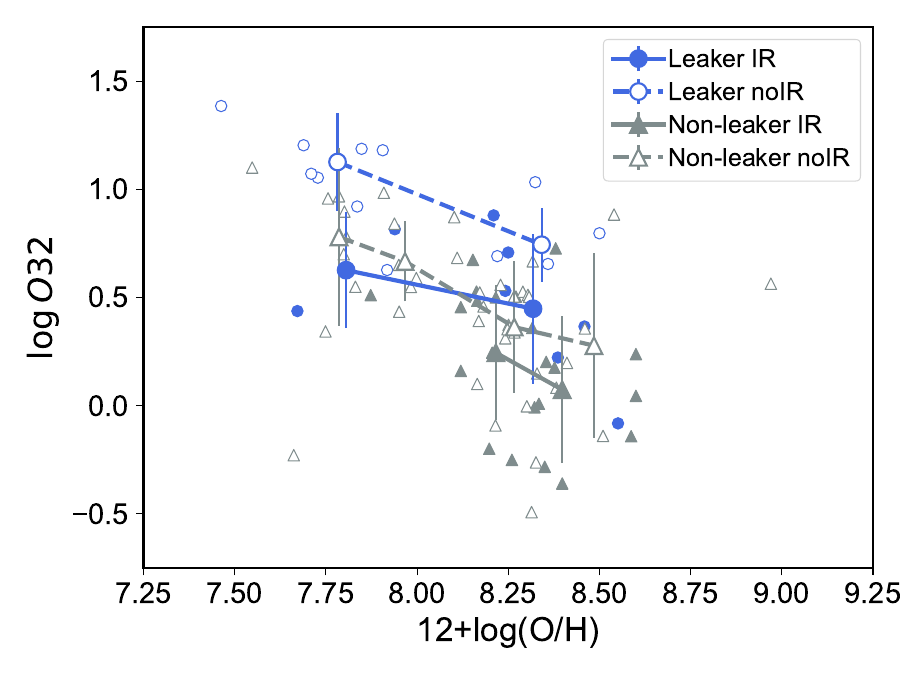}
    \caption{O32 as a function of gas-phase metallicity, $12+\log({\rm O/H})$ for the strong leakers (circles) and non-leakers (triangles). IR-detected galaxies are shown as filled symbols, while IR-undetected galaxies are shown as open symbols. For each subsample, galaxies are further divided into metallicity bins, and the large symbols and error bars indicate the median and standard deviation of O32 in each bin. The IR-undetected leakers show higher O32 than the other subsamples across the overlapping metallicity range, suggesting that their elevated O32 is not solely driven by their lower gas-phase metallicity.}
    \label{fig:metal_o32}
\end{figure}

On the other hand, the IR-undetected strong leakers have systematically higher $f_{\rm esc}$ and may represent a more extreme escape regime. All of our IR-undetected strong leakers have $A_{\rm r}<3$, apart from one object with a complex triple-peaked Ly$\alpha$ profile for which this diagnostic is not applicable, broadly consistent with the density-bounded dominated regime proposed by \citet{kakiichi2021}. \citet{gazagnes2020} also find that the fraction of low column density channels increases in higher $f_\mathrm{esc}$ LyC leakers. \citet{ramambason2020} introduced a two-zone photoionization model to explain the line ratios in LyC leakers. They found that the contribution from highly ionized, density-bounded regions becomes increasingly important toward higher $f_{\rm esc}$. This provides a possible physical interpretation of the difference between our two subsamples of strong LyC leakers, including the higher $f_{\rm esc}$ and O32 values of the IR-undetected leakers. We also note that the higher O32 values of the IR-undetected leakers may partly reflect their lower metallicity; however, they still tend to have higher O32 than non-leakers over the same metallicity range (Figure \ref{fig:metal_o32}), suggesting that metallicity alone cannot account for their enhanced ionization state and that these IR-undetected strong leakers are likely to have a higher ionization parameter or a more intense ionizing radiation field. This interpretation of the IR-undetected strong leakers should nevertheless be treated with caution because the absence of a WISE detection does not necessarily imply the absence of dust emission below the WISE sensitivity limit. The dust distribution further complicates the picture \citep[e.g.,][]{zakrisson_inoue_jensen2013}. Current data also do not distinguish between a very porous, anisotropic system dominated by density-bound channels and an isotropic, density-bounded one. Future JWST observations may provide useful constraints on the ISM structure and help clarify the scenario.

\subsection{Dust attenuation in LyC leakers}
\label{subsec:dust}

Dust has long been recognized as an important source of uncertainty in studies of LyC escape \citep[e.g.,][]{mathis1971,zakrisson_inoue_jensen2013,chisholm2018,kakiichi2021}. In particular, it remains unclear to what extent dust can absorb LyC photons at wavelengths below 912 {\AA} and how closely dust is spatially associated with the gas through which LyC photons propagate. The WISE observations provide additional information on the dust content and attenuation of LyC leakers, although the mid-infrared data are not as powerful as far-infrared data in constraining the dust properties. As mentioned in Section \ref{subsec:physprop}, the $E(B-V)$ values derived from SDSS emission-line measurements for IR-detected leakers are lower than those for non-leakers despite their similar dust luminosities and masses. Here, we further compare the $E(B-V)$ with the IR/UV ratio inferred from broadband observations for the IR-detected subsamples to explore what these results may reveal about dust attenuation and geometry in LyC leakers.

\begin{figure}
    \centering
    \includegraphics[width=\linewidth]{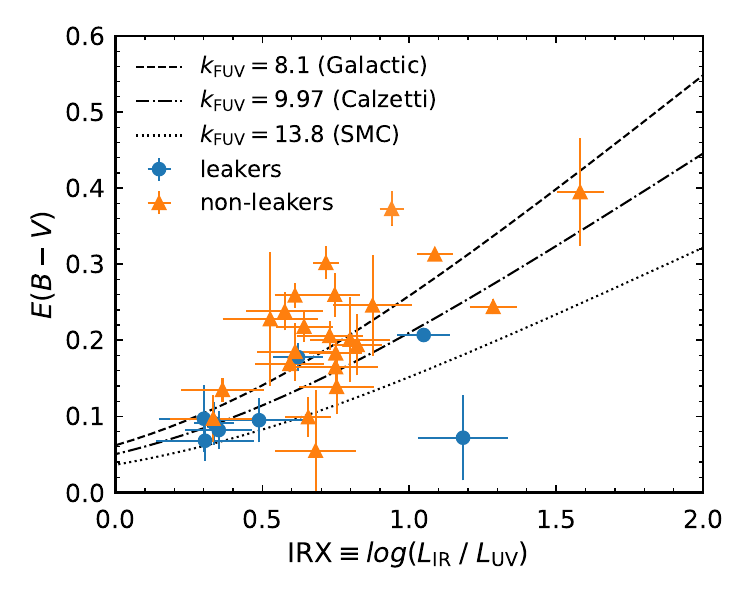}
    \caption{Comparison of $E(B-V)$ derived from the SDSS spectra and $\log L_\mathrm{IR}/L_\mathrm{UV}$ (IRX) for the IR-detected leakers (dots) and non-leakers (triangles) in our sample. The relations expected from theoretical attenuation curves are overplotted.  }
    \label{fig:irx_ebv}
\end{figure}


Figure \ref{fig:irx_ebv} compares $E(B-V)$ with $\log L_\mathrm{IR}/L_\mathrm{UV}$ (IRX) for the IR-detected galaxies. $L_\mathrm{IR}$ is the $L_\mathrm{dust}$ derived from CIGALE. $L_\mathrm{UV}$ is derived from the FUV flux of GALEX data ($\lambda_\mathrm{UV}\approx1530\AA$). We overplot the relation under three dust attenuation curves using the formula $\mathrm{IRX}=C(10^{0.4 k_{UV}E(B-V)} - 1.0)$. The $k_\mathrm{UV}$ depends on the extinction curve $k_\lambda$. The factor $C$ accounts for the ratio between the total absorbed stellar luminosity and the absorbed luminosity at the UV wavelength. Based on the range of best-fit intrinsic SEDs for our sample, we adopt $C=1.7$. The LyC leakers tend to have lower $E(B-V)$ than the non-leakers, despite having broadly similar IRX values. Therefore, the leakers seem to lie closer to the relation expected for a steeper attenuation curve than non-leakers. This could potentially indicate different dust grain properties in the LyC leakers. However, the large scatter prevents us from identifying a clear difference in the IRX{--}$E(B-V)$ relation for the leakers and non-leakers. 
Moreover, here IRX is derived from CIGALE under an energy-balance framework and therefore depends on the adopted attenuation and dust emission prescriptions, making it difficult to use the current IRX{--}$E(B-V)$ relation to distinguish between different attenuation curves and thus dust grain properties. The lower $E(B-V)$ at comparable IRX may instead reflect differences in the spatial regions probed by the two quantities: $E(B-V)$ from emission lines traces the dust column toward the ionized gas, whereas IRX provides a more global constraint on dust-absorbed stellar emission. The lower nebular reddening of the leakers is therefore qualitatively consistent with a porous ISM in which LyC photons preferentially escape through low-column-density channels.

The similar dust masses but lower $E(B-V)$ of the IR-detected leakers should also not be interpreted straightforwardly as evidence for a more extended dust distribution. In an inhomogeneous ISM, the observed $E(B-V)$ can be strongly affected by the covering fraction and the particular sightlines to the emitting regions, so a lower effective dust column does not necessarily imply a globally more diffuse dust distribution. Instead, it may reflect the presence of relatively transparent channels within an otherwise substantial dust reservoir. More sensitive and spatially resolved IR observations will be needed to distinguish between these possibilities and to constrain the relation between dust, gas, and LyC escape.

\subsection{Connection between low-z and high-z LyC leakers}

\begin{figure*}
    \centering
    \includegraphics[width=0.75\linewidth]{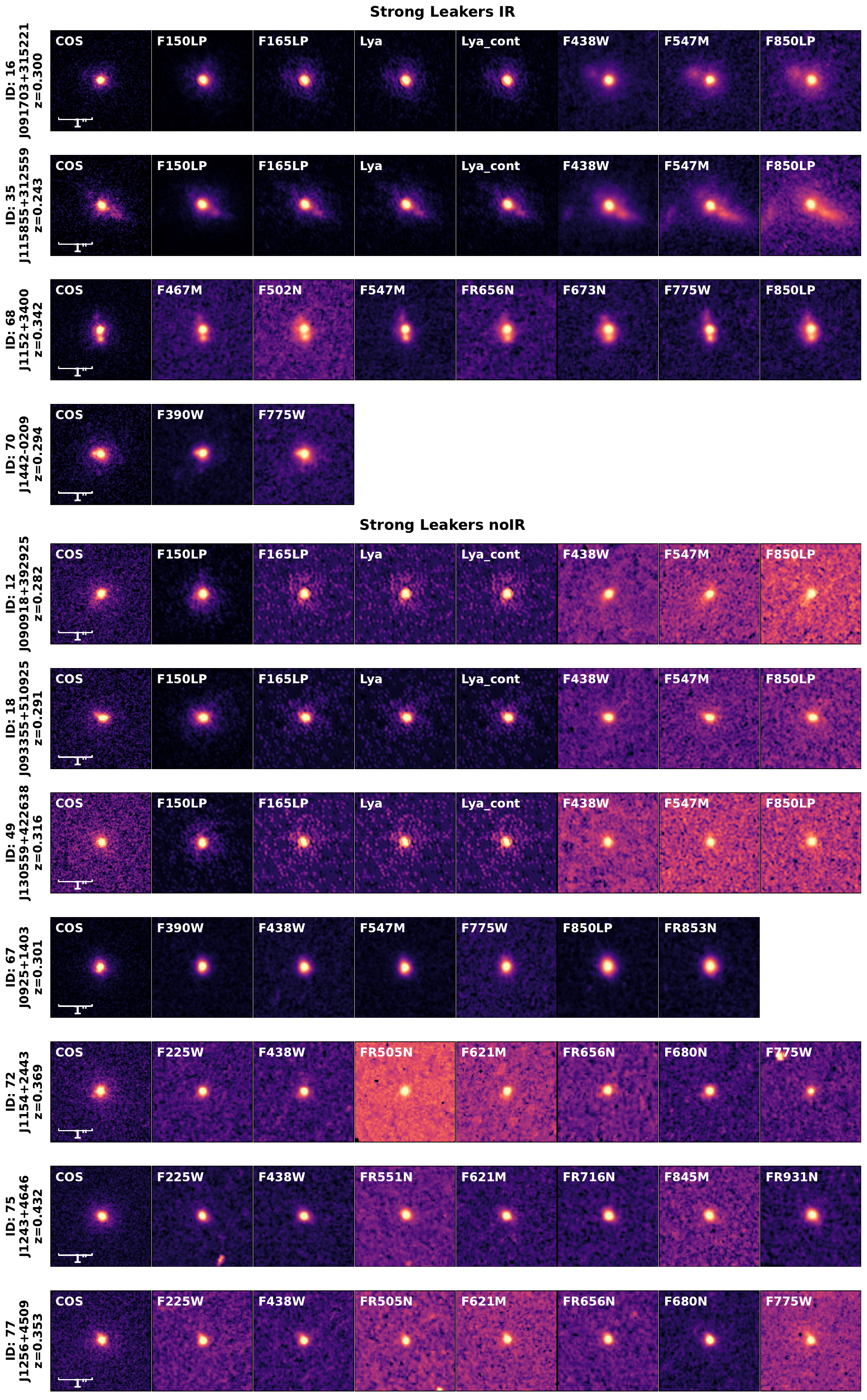}
    \caption{HST images for strong LyC leakers ($f_\mathrm{esc}>5\%$). Top: IR-detected strong leakers. Bottom: IR-undetected strong leakers.}
    \label{fig:images}
\end{figure*}

The IR-faint and IR-bright systems identified in our sample may respectively resemble the compact starburst and merger-associated LCEs discussed in recent high-redshift ($z>3$) studies \citep{zhu2025}. While compact starbursts may favor density-bounded escape, more extended or interacting systems could naturally produce fragmented ISM geometries that lead to anisotropic LyC leakage. The role of galaxy mergers in facilitating LyC escape remains a subject of active debate \citep[e.g.,][]{yuan2024, lereste2025merger, mascia2025,wang2025}.

In our sample, 11 strong leakers have been observed with HST ACS and UVIS. Figure~\ref{fig:images} presents the multiwavelength stamps of these galaxies. We also include Ly$\alpha$ and Ly$\alpha$ continuum images from \citet{lereste2025sample}. The IR-detected leakers exhibit extended or multi-component morphologies, showing features suggestive of galaxy interactions or mergers, whereas the IR-undetected leakers are generally more compact and do not show obvious multi-clump structures. 

To quantify this difference, we measure the rest-frame optical half-light radius ($r_{50}^\mathrm{opt}$) using the F850LP or F775W images when available; for J1243+4646, we use F845M due to the lack of both bands. The average $r_{50}^\mathrm{opt}$ of the IR-detected strong leakers ($0.89 \pm 0.28$ kpc) is nearly twice that of the IR-undetected ones ($0.48 \pm 0.05$ kpc), supporting the visual impression that the IR-detected systems are typically more extended. We note that the multi-wavelength coverage is important for identifying the extended features, as different bands trace different structural components of galaxies. For example, in J091703+315221, the secondary component is only visible at optical wavelengths (F438W, F547M, and F850LP). Therefore, optical images provide complementary information on the underlying stellar distribution of these galaxies that may not be fully captured by UV images alone. This wavelength dependence may explain why the COS-based UV sizes ($r_{50}^{\mathrm{COS}}$ in Table \ref{tab:IR_comparison}) are similar between the IR-detected and IR-undetected subsamples.

Although the current sample remains limited, these results suggest that IR-detected and IR-undetected strong leakers may be associated with different structural conditions related to LyC escape. Their structural diversity—ranging from compact systems to interacting galaxies—appears to be similar to that seen in high‑redshift LCEs.

In the local sample, IR‑undetected strong leakers dominate the total ionizing budget, contributing roughly 67\% of the emitted LyC photons, while IR‑detected strong leakers provide a smaller but non‑negligible fraction ($\sim$20\%). Given the complex selection function of the sample, these fractions should be interpreted with caution. 
Recent studies have reported an increasing number of dusty galaxies at the epoch of reionization, although their contribution to cosmic reionization remains unknown. Considering that the dust reservoirs in our IR-detected strong leakers ($M_\mathrm{dust} \approx 4.0 \times 10^7 M_\odot$) are comparable to those inferred for dusty galaxies at the EoR \citep[e.g.,][]{watson2015, fudamoto2021}, such IR-bright systems may provide useful analogs for understanding the role of dust-rich galaxies in cosmic reionization under a potentially dustier early-universe environment \citep[e.g.,][]{sun2026}.

\section{Summary}
\label{sec:summary}

We investigate the IR properties of a local sample of 89 star-forming galaxies, including 20 strong LyC leakers, compiled from the LzLCS and archival data presented by \citet{Flury2022}. We divide the sample into four subsamples according to IR detection and LyC leakage strength and compare their properties. Our findings include:
\begin{enumerate}

    \item Strong LyC leakers are present in both IR-undetected and IR-detected systems. The IR-detected leakers maintain significant LyC escape fractions despite their higher IR luminosities.
    
    \item  IR-undetected strong leakers are characterized by the highest O32, bluest $\beta_{1200}$ slopes, lowest metallicity, and the lowest $E(B-V)$ among all subsamples. In contrast, the global properties of IR-detected strong leakers, including O32, $\beta_{1200}$, and metallicity, are similar to those of non-leakers. The IR-detected leakers exhibit low $E(B-V)$ values despite their high IR luminosities, indicating that the optical and IR components in these systems likely arise from different sites.

    \item IR-detected strong leakers are dominated by structured or interacting systems and have a larger optical size than the IR-undetected leakers on average.
\end{enumerate}

The different properties suggest that IR-detected and IR-undetected strong leakers may represent different physical pathways for LyC escape. The IR-detected strong leakers likely leak LyC through low-density channels in dusty environments, whereas the IR-undetected strong leakers are more consistent with highly ionized, density-bounded systems. The Ly$\alpha$ and radio data are compatible with this scenario, though the sample remains small.

Our results show that strong LyC escape can arise under a broad range of physical conditions. This diversity is in line with the emerging picture from high-redshift studies that LyC leakage can occur in both compact starbursts and more extended or interacting systems \citep[e.g.,][]{marques-chaves2021,marques-chaves2024,kerutt2024,zhu2024,zhu2025}. Further investigation of the diversity of LyC escape conditions will require larger LCE samples and higher-resolution, higher-sensitivity observations spanning the UV-to-IR wavelength range.

\begin{acknowledgements}
    
We would like to thank the anonymous referee for helpful comments that have improved the quality of the manuscript. This work is supported by the China Survey Space Telescope (CSST) project dedicated scientific research funds under grant Nos. CMS-CSST-2025-A06, CMS-CSST-2025-A10, CMS-CSST-2025-A17, and CMS-CSST-2025-A18. ZYZ acknowledges the supports by the Shanghai Leading Talent Program of Eastern Talent Plan (LJ2025051) and the China-Chile Joint Research Fund (CCJRF No. 1906).

The {\it WISE} data used in this work are available from the Infrared Science Archive (IRSA) at IPAC: \dataset[10.26131/IRSA1]{http://dx.doi.org/10.26131/IRSA1}.HST data presented in this paper were obtained from the Mikulski Archive for Space Telescopes (MAST) at the Space Telescope Science Institute. The specific observations analyzed can be accessed via \dataset[https://doi.org/10.17909/p743-fq71]{https://doi.org/10.17909/p743-fq71}. STScI is operated by the Association of Universities for Research in Astronomy, Inc., under NASA contract NAS5–26555. Support to MAST for these data is provided by the NASA Office of Space Science via grant NAG5–7584 and by other grants and contracts.

Funding for the Sloan Digital Sky Survey IV has been provided by the Alfred P. Sloan Foundation, the U.S. Department of Energy Office of Science, and the Participating Institutions. SDSS-IV acknowledges
support and resources from the Center for High-Performance Computing at
the University of Utah. The SDSS web site is www.sdss.org.

SDSS-IV is managed by the Astrophysical Research Consortium for the 
Participating Institutions of the SDSS Collaboration including the 
Brazilian Participation Group, the Carnegie Institution for Science, 
Carnegie Mellon University, the Chilean Participation Group, the French Participation Group, Harvard-Smithsonian Center for Astrophysics, 
Instituto de Astrof\'isica de Canarias, The Johns Hopkins University, Kavli Institute for the Physics and Mathematics of the Universe (IPMU) / 
University of Tokyo, the Korean Participation Group, Lawrence Berkeley National Laboratory, 
Leibniz Institut f\"ur Astrophysik Potsdam (AIP),  
Max-Planck-Institut f\"ur Astronomie (MPIA Heidelberg), 
Max-Planck-Institut f\"ur Astrophysik (MPA Garching), 
Max-Planck-Institut f\"ur Extraterrestrische Physik (MPE), 
National Astronomical Observatories of China, New Mexico State University, 
New York University, University of Notre Dame, 
Observat\'ario Nacional / MCTI, The Ohio State University, 
Pennsylvania State University, Shanghai Astronomical Observatory, 
United Kingdom Participation Group,
Universidad Nacional Aut\'onoma de M\'exico, University of Arizona, 
University of Colorado Boulder, University of Oxford, University of Portsmouth, 
University of Utah, University of Virginia, University of Washington, University of Wisconsin, 
Vanderbilt University, and Yale University.
\end{acknowledgements}

\bibliography{lyclowz_bib}{}
\bibliographystyle{aasjournal}

\end{document}